\documentclass[twocolumn]{svjour3}          
\smartqed  
\usepackage{graphicx}
\usepackage{amsmath,subfig}
\begin{document}

\title{Pair correlation function of charge-stabilized colloidal systems under sheared conditions}


\author{Luca Banetta         \and
        Alessio Zaccone 
}


\institute{L. Banetta \at
              Department of Chemical Engineering and Biotechnology, University of Cambridge, Cambridge, CB3 0AS, United Kingdom \\
              \email{lb725@cam.ac.uk}     \\                      
              \and
              A. Zaccone \at
              Cavendish Laboratory, University of Cambridge, JJ Thomson Avenue, CB3 0HE Cambridge, U.K.\\
              \emph{Present address: Department of Physics ``A. Pontremoli", University of Milan, via Celoria 16, 20133 Milan, Italy}}

\date{Received: date / Accepted: date}

\maketitle

\begin{abstract}
	The pair correlation function of charge stabilized colloidal particles under strongly sheared conditions is studied using the analytical intermediate asymptotics method recently developed in [L. Banetta and A. Zaccone, Phys. Rev. E 99, 052606 (2019)] to solve the steady-state Smoluchowski equation for medium to high values of the P\'eclet number; the analytical theory works for dilute conditions. A rich physical behaviour is unveiled for the pair correlation function of colloids interacting via the repulsive Yukawa (or Debye-H\"uckel) potential, in both the extensional and compressional sectors of the solid angle. In the compression sector, a peak near contact is due to the advecting action of the flow and decreases upon increasing the coupling strength parameter $\Gamma$ of the Yukawa potential. Upon increasing the screening (Debye) length $\kappa^{-1}$, a secondary peak shows up, at a larger separation distance, slightly less than the Debye length. While this secondary peak grows, the primary peak near contact decreases. The secondary peak is attributed to the competition between the advecting (attractive-like) action of the flow in the compressions sector, and the repulsion due to the electrostatics. 
	In the extensional sectors, a depletion layer (where the pair-correlation function is identically zero) near contact is predicted, the width of which increases upon increasing either $\Gamma$ or $\kappa^{-1}$. 
\end{abstract}

\section{Introduction}
The microstructure, that is the spatial arrangement, of interacting colloidal particles embedded in a viscous liquid is an important problem in physical chemistry with many applications ranging from emulsions, polymerization processes in aqueous phase, to atmospheric science and consumers products. 
The single quantity which provides all the information about the microstructure of a colloidal suspension is 
the pair correlation function $g(\textbf{r},t)$ that is the probability to find  $N$ particles in positions $\textbf{r} = (\textbf{r}_1, ... , \textbf{r}_N)$ at time $t$ ~\cite{Larsen,Hansen}.
This is the solution to the stochastic N-body Smoluchowski equation:
\begin{equation}
\dfrac{\partial }{\partial t} g(\textbf{r},t)  = \sum_{i,j=1}^\text{N} \nabla_{\textbf{r}_i} \cdot \textbf{D}^{\text{Br}} \cdot \biggl[ - \beta \textbf{\textit{K}}_i^{\text{int}} + \nabla_{\textbf{r}_j} g(\textbf{r},t) \biggr],
\label{Smoluchowski_Brownian}
\end{equation}
where $\textbf{D}^{\text{Br}}$ is the microscopic diffusion matrix which describes the influence of the medium on particles moving under Brownian dynamics , $\textbf{\textit{K}}_i^\text{int}$ is the force acting on the $i$-th particle due to the pair-wise interactions with the other $N-1$ particles, and $\beta = 1/k_\text{B}T$ with $k_\text{B}$ being the Boltzmann constant and $T$ the absolute temperature.\\
Equation (\ref{Smoluchowski_Brownian}) has been adopted to study the influence of Brownian motion and inter-particle interactions on the micro-structure of colloidal suspensions \cite{Dhont_Book} or dusty plasmas \cite{Fortov}. However, considerable less information and understanding are available for colloidal systems that are subject to a laminar shear flow, in spite of the great of importance of this situation for industrial applications~\cite{Guido}, crystallization phenomena~\cite{Mura}, complex plasmas~\cite{Lowen} and atmospheric science~\cite{Falkovich}.\\

Earlier numerical work on this problem has been focused on scenarios where the Reynolds number is sufficiently low that possible inertial effects acting on the particles are negligible; the systems which fulfill this condition can be described numerically using \textit{Stokesian dynamics}, which accounts for the role of hydrodynamic interactions between the particles. The hydrodynamic interactions, in turn, arise due to the (incompressible) liquid medium being displaced by the particles motion.\\
At the level of theory, Eq.(\ref{Smoluchowski_Brownian}) can be modified to account for contributions due to the presence of a flow field \cite{Dhont_Book}.
In general, it is important to be aware that the microstructure of a colloidal suspension is dependent on two parameters: the volume fraction $\phi$ occupied by the particles of radius $a$, $\phi = 4/3 \pi a^3 N$, and the relative importance of Brownian- and shear-induced effects, which is described by the P\'eclet number \cite{MorrisBrady}
\begin{equation}
\text{Pe} = \dfrac{a^2 \dot{\gamma}}{D_0} = \dfrac{6 \pi \eta \dot{\gamma} a^3}{k_B T}.
\label{Peclet_number}
\end{equation}
The starting point has been the pioneering paper by Batchelor and Green \cite{BatchelorGreen} who derived an analytic solution of the two-body the Smoluchowski equation (i.e. Eq. (1) with $N=2$) under shear flow for hard spheres. The pair correlation function (pcf), i.e. the probability of finding a particle at a certain position $\textbf{r}$ with respect to a reference particle placed at the origin of the spherical frame, was evaluated for the limiting case of infinite P\'eclet number.\\

Later theoretical work \cite{Blawdz,MorrisBrady} evidenced the characteristic shear-induced distortion of the pcf, with an asymmetric distribution of the probability of finding particles around the reference particle, in the solid angle. If we are in a situation where the shear flow pushes the particles towards each other (compression sectors), then the pcf features an accumulation peak whose magnitude depends on the P\'eclet number. On the other hand, in the sectors of solid angle where the shear tends to separate the particles from each other, the pcf takes values which are much lower next to the surface of the reference particle.\\
The distortion of the pcf at finite Pe has also been proved by computational simulations of colloidal suspensions using Stokesian dynamics (SD) \cite{BradyBossis,Morris}: even at high packing fractions $\phi$, the microstructure presents an accumulation peak in the compression sectors and lower values in the extensional ones.
In recent years, new analytical formulations for the two-body Smoluchowski equation have been derived including many body effects to describe the microstructure of more concentrated systems for both hard-spheres \cite{NazockdastMorris} and interacting soft spheres \cite{NazockdastMorrisSoftSpheres}, but the solution of the equation in spherical coordinates is fully numerical. 
As a consequence, an analytical framework which describes the micro-structure of complex interacting particles under shear flow is still lacking.\\  

Recently, a theory based on intermediate asymptotics expansions has been developed,  which analytically describes the micro-structure of a dilute suspension of particles. The work has been validated by comparison with numerical simulation data of hard spheres from Stokesian dynamics \cite{Morris} and it has been found out that the predictions are valid for semi-dilute conditions ($\phi$ up to 0.2) under  strongly simple sheared conditions \cite{Banetta}. The reason for this is a cancellation of errors between the neglect of the tangential contribution to the lubrication forces acting on Brownian motion and the absence of many-body interactions. The theory has then been used to obtain the first prediction of the pair correlation function of attractive Lennard-Jones particles in shear flow.
Here, we extend this methodology to the description of the microstructure of charge stabilized colloids interacting through the screened Coulomb (Yukawa or Debye-H\"uckel) potential, under simple shear flow.\\
\section{Model}
We start from considering the steady state two-body ($N=2$) limit of Eq.(\ref{Smoluchowski_Brownian}), which can describe dilute and semi-dilute suspensions up to $\phi \sim 0.20$:
\begin{equation}
\nabla \cdot \textbf{D}^{\text{Br}} \cdot \biggl[ -  \beta \textbf{\textit{K}}^{\text{int}} g(\textbf{r}) + \nabla g(\textbf{r}) \biggr] = 0.
\label{Smoluchowski_Steady_State}
\end{equation}
It is important to notice that we have written Eq.(\ref{Smoluchowski_Steady_State}) as a function of \textbf{r} = $\textbf{r}_2 - \textbf{r}_1 = (r,\theta,\phi)$, the relative position of a second particle with respect to the \textit{reference} particle placed at the center of the spherical frame.
\subsection{Brownian contributions}
Before moving on to considering the contribution from shear flow, it is important to define each term in Eq.(\ref{Smoluchowski_Steady_State}).
To model the microscospic diffusion matrix $\textbf{D}^\text{Br}$, we need to consider the effect of the presence of a viscous medium between the particles.
If the particles get closer and closer to each other, the squeezing of the fluid between them causes a repulsive effect called \textit{lubrication force} which opposes their further approach \cite{BradyBossis,Brenner}; we will consider their effect by adopting the following constitutive equation for $\textbf{D}^\text{Br}$:
\begin{gather}
\textbf{D}^{\text{Br}} = 2 D_0 
\begin{bmatrix} G(r) & 0 & 0 \\ 0 & H(r) & 0 \\ 0 & 0 & H(r) \end{bmatrix} = 2 D_0 \ \underline{\textbf{D}}^\text{Br},
\label{Brownian_Induced_Diffusion_Matrix}
\end{gather}
where $D_0$ is the diffusion coefficient of an isolated particle and $G(r)$ a parametrized function which approximates the rigorous solution for the lubrication force component of the hydrodynamic interactions \cite{Honig,Banetta} along the line of centres, meanwhile $H(r)$ is its equivalent relative to the tangential directions with respect to the motion of the colloids; in the calculations we will consider only the contribution of the lubrication forces along the radial directions, which means $H(r) = 0$.\\
Finally, the conservative interaction force $\textbf{\textit{K}}_\text{int}$ is given by $\textbf{\textit{K}}^\text{int} = (- \nabla U(r), 0 ,0) $, where $U(r)$ is the interaction potential between two particles.
\begin{figure}[ht]
	\centering
	\includegraphics[width = 0.7 \linewidth]{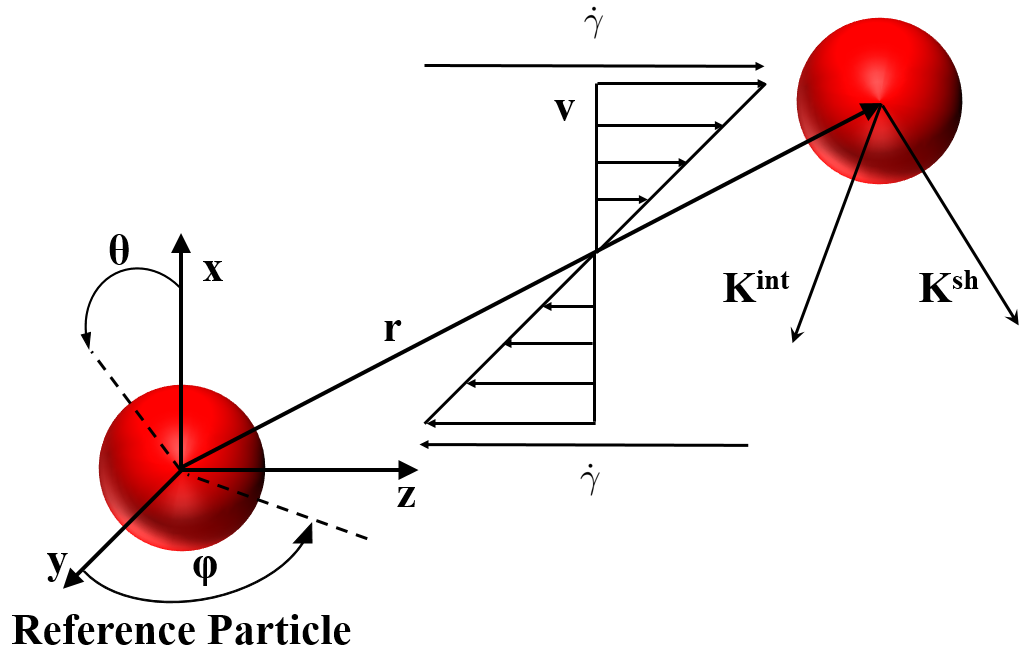}
	\caption{Schematic illustration of a pair of interacting particles subject to a simple shear flow where \textbf{v} = (0,0,$\dot{\gamma}$x); the spherical reference frame has been taken as described in \cite{Adler}.}
	\label{fig:physical_system}
\end{figure}
\subsection{Shear-induced contributions}
We can model the influence of an external flow field through the introduction of an extra term in Eq.(\ref{Smoluchowski_Steady_State}) \cite{Dhont_Book}:
\begin{equation}
0 = \nabla \cdot \textbf{D}^{\text{Br}} \cdot \biggl[ - \beta \textbf{\textit{K}}^\text{int} g(\textbf{r}) + \nabla g(\textbf{r}) \biggr] + \nabla \cdot  \textbf{D}^{\text{sh}} \cdot \biggl[-\beta \textbf{\textit{K}}^\text{sh}g(\textbf{r})\biggr], 
\label{Smoluchowski_Shear_Steady_State}
\end{equation}
where $\textbf{D}^{\text{sh}}$ is the microscopic diffusion matrix relative to to the disturbance of the flow field around the particles due to the application of a shear stress. Here, $\textbf{\textit{K}}^\text{sh}$ is the (non-conservative) drag force according to Fig.\ref{fig:physical_system}, which in our two-body description is directly proportional to the relative velocity \textbf{v}(\textbf{r}):
\begin{equation}
\textbf{\textit{K}}^{\text{sh}} = \zeta \textbf{v}(\textbf{r}).
\label{shear_force}
\end{equation}
In particular, $\zeta = 6 \pi \eta a$ is the Stokes drag coefficient, which depends on the particle radius $a$ and the viscosity of the liquid medium $\eta$.\\
We will model $\textbf{D}^{\text{sh}}$ in the simplest way possible and consider the presence of the effects of the reflective flow from one particle to the other in the definition of the relative velocity:
\begin{equation}
\textbf{D}^{\text{sh}} = D_0 \textbf{I},
\end{equation}
where $D_0$ is the mutual diffusion coefficient between the particles; from the resolution of the creeping flow equations, it is possible to introduce $\textbf{v}(\textbf{r})$ as \cite{Lin,Adler}
\begin{equation}
\begin{cases}
v_r = \dot{\gamma} r (1-A(r))\sin^2{\theta} \sin\phi \cos\phi; \\
v_{\theta} = \dot{\gamma} r (1- B(r)) \sin{\theta} \cos{\theta}\sin\phi \cos\phi;\\
v_{\phi} = \dot{\gamma} r \sin\theta\biggl( \cos^2\phi - \dfrac{B(r)}{2} \cos(2 \phi) \biggr).
\label{relative_velocity_shear_flow}
\end{cases}
\end{equation}
It is important to highlight that the relative velocity between the particles is the superposition of two effects: one is due to the motion of the fluid because of the applied shear, the second is a reflected flow from one particle to the other, a distortion of the flow field around one particle due to the presence of the other. The latter contribution to \textbf{v}(\textbf{r}) is represented by $A(r)$ and $B(r)$, hydrodynamic functions derived from a rigorous resolution of the Stokes equations for incompressible fluids ~\cite{Lin,Batchelor}; more information about these terms can be found in the Appendix B.\\
The Smoluchowski equation for the pcf $g(\textbf{r})$ then becomes
\begin{multline}
2 D_0 \nabla \cdot \underline{\textbf{D}}^\text{Br} \cdot \biggl( - \beta \textbf{K}^\text{int}  g(\textbf{r}) + \nabla g(\textbf{r}) \biggr) + \\ - D_0 \nabla \cdot \biggl(\beta \zeta \textbf{v}(\textbf{r}) g(\textbf{r})\biggr) = 0. 
\label{Smoluchowski_Shear_Steady_State_1}
\end{multline}
which will be the starting point for the mathematical evaluation of the pcf $g(\textbf{r})$ in the following.
\section{Formulation of the mathematical problem}
First, we make Eq.(\ref{Smoluchowski_Shear_Steady_State_1}) dimensionless through:
\begin{equation}
\begin{cases}
\tilde{\nabla} = \sigma \nabla, \\
\tilde{\textbf{K}}^\text{int} = \beta \textbf{K}^\text{int}, \\
\end{cases}
\end{equation}
where $\sigma = 2 a$  is the hard-core particle diameter.\\
The velocity $\textbf{v(r)}$ can be expressed as:
\begin{equation}
\textbf{v(r)} =\dot{\gamma} \sigma \tilde{\textbf{v}}(\tilde{\textbf{r}}).
\end{equation}
The same can be done with the interaction potential which becomes $\tilde{U}(\tilde{r})$ and Eq.(\ref{Smoluchowski_Shear_Steady_State_1}) can be rewritten as
\begin{multline}
2 \ \tilde{\nabla} \cdot \underline{\tilde{\textbf{D}}}^\text{Br} \cdot \biggl( - \tilde{\textbf{K}}^\text{int} g(\tilde{\textbf{r}}) + \tilde{\nabla} g(\tilde{\textbf{r}}) \biggr)+ \\  - \tilde{\nabla} \cdot \biggl(4 \dfrac{6 \pi \eta \dot{\gamma} a^3 }{k_B T} \tilde{\textbf{v}}(\tilde{\textbf{r}}) g(\tilde{\textbf{r}})\biggr) = 0. 
\label{Smoluchowski_dimensionless}
\end{multline}
Recalling the P\'eclet number already introduced in Eq.(\ref{Peclet_number}) we can write
\begin{multline}
\tilde{\nabla} \cdot \underline{\tilde{\textbf{D}}}^\text{Br} \cdot \biggl(- \tilde{\textbf{K}}^\text{int} g(\tilde{\textbf{r}}) + \tilde{\nabla} g(\tilde{\textbf{r}}) \biggr) + \\  - 2 \text{Pe} \tilde{\nabla} \cdot \biggl( \tilde{\textbf{v}}(\tilde{\textbf{r}}) g(\tilde{\textbf{r}}) \biggr) = 0. 
\label{Smoluchowski_dimensionless_2}
\end{multline}
Equation (\ref{Smoluchowski_dimensionless_2}) is to be solved perturbatively. 
A perturbative method is based on the introduction of a small perturbation parameter $\epsilon$, by definition much smaller than unity, which simplifies the analytical treatment of the partial differential equation (PDE) of interest \cite{BenderOrszag,Hinch,VanDyke}.
Focusing on situations where the effect of shear flow is substantial, we fix:
\begin{equation}
\epsilon = \dfrac{1}{\text{Pe}}.
\label{Perturbationparameter}
\end{equation}
Applying Eq.(\ref{Perturbationparameter}) to Eq.(\ref{Smoluchowski_dimensionless_2}) we obtain:
\begin{multline}
\epsilon \biggl[ \tilde{\nabla} \cdot \underline{\tilde{\textbf{D}}}^\text{Br} \cdot \biggl( - \tilde{\textbf{K}}^\text{int} g(\tilde{\textbf{r}}) + \tilde{\nabla} g(\tilde{\textbf{r}}) \biggr) \biggr]+ \\  - 2 \ \tilde{\nabla} \cdot \biggl( \tilde{\textbf{v}}(\tilde{\textbf{r}}) g(\tilde{\textbf{r}}) \biggr) = 0. 
\label{Smoluchowski_epsilon}
\end{multline}
Starting from Eq.(\ref{Smoluchowski_epsilon}), we apply the linearity of the divergence operators obtaining:
\begin{multline}
\epsilon  \biggl[ \tilde{\nabla} \cdot \biggl( \underline{\tilde{\textbf{D}}}^\text{Br} \cdot \tilde{\nabla} g(\tilde{\textbf{r}}) \biggr) - \tilde{\nabla} \cdot \biggl( \underline{\tilde{\textbf{D}}}^\text{Br} \cdot   \tilde{\textbf{K}}^\text{int}\biggr) g(\tilde{\textbf{r}})  \biggr] + \\ - 2 \biggl( \tilde{\textbf{v}} \cdot \tilde{\nabla} g(\tilde{\textbf{r}}) + g(\tilde{\textbf{r}}) \tilde{\nabla}\cdot\tilde{\textbf{v}} \biggr) = 0.
\label{Smoluchowski_epsilon_1}
\end{multline}
It is important to remember that $\tilde{\textbf{v}}$ is the relative velocity between the particles, so its divergence can assume not null values, even if the fluid is incompressible.
Next, we introduce a useful approximation that was proposed in \cite{ZacconePRE2009} (see also ~\cite{NazockdastMorris}) in order to make the 3D problem analytically solvable. The approximation consists in applying an angular average, denoted as $\langle \cdots \rangle$, over a certain portion of solid angle to Eq.(\ref{Smoluchowski_epsilon_1}).
\begin{multline}
\epsilon  \biggl[ \tilde{\nabla} \cdot \biggl( \underline{\tilde{\textbf{D}}}^\text{Br} \cdot \tilde{\nabla} \langle g(\tilde{\textbf{r}}) \rangle \biggr) - \tilde{\nabla} \cdot \underline{\tilde{\textbf{D}}}^\text{Br} \cdot \biggl(  \tilde{\textbf{K}}^\text{int} \langle g(\tilde{\textbf{r}}) \rangle \biggr) \biggr] + \\ - 2 \biggl( \langle \tilde{\textbf{v}} \cdot \tilde{\nabla} g(\tilde{\textbf{r}}) \rangle + \langle g(\tilde{\textbf{r}}) \tilde{\nabla}\cdot\tilde{\textbf{v}} \rangle \biggr) = 0.
\label{Smoluchowski_epsilon_2}
\end{multline}
Since we have neglected the tangential contribution of the lubrication forces acting on the Brownian motion, we can apply the angular average directly on the pcf when it comes to the section of Eq.(\ref{Smoluchowski_epsilon_2}) related to the Brownian contribution; a more detailed explanation about it has been proposed in Appendix A.\\
This procedure can be taken, for example, over the extensional sectors only, or over the compression sectors, thus leading to the pcf separately averaged in extension and in compression (see the Appendix B). 
The result is the following spherically-averaged solution $g(\tilde{r})$ over a certain region (either extensional or compressional) which now depends on the radial coordinate only:
\begin{multline}
\epsilon  \biggl[ \tilde{\nabla} \cdot \biggl( \underline{\tilde{\textbf{D}}}^\text{Br} \cdot \tilde{\nabla} g(\tilde{r})  \biggr) - \tilde{\nabla} \cdot \underline{\tilde{\textbf{D}}}^\text{Br} \cdot \biggl(  \tilde{\textbf{K}}^\text{int}  g(\tilde{r}) \biggr) \biggr] + \\ - 2 \biggl( \langle \tilde{\textbf{v}} \cdot \tilde{\nabla} g(\tilde{\textbf{r}}) \rangle + \langle g(\tilde{\textbf{r}}) \tilde{\nabla}\cdot\tilde{\textbf{v}} \rangle \biggr) = 0.
\label{Smoluchowski_epsilon_3}
\end{multline}
Moreover, we use a weak-coupling approximation between flow field and particle concentration field also introduced in \cite{ZacconePRE2009}; we suppose that the velocity and the pair correlation function are weakly correlated, so that:
\begin{equation}
\langle \tilde{\textbf{v}} \cdot \tilde{\nabla} g(\tilde{\textbf{r}}) \rangle + \langle g(\tilde{\textbf{r}}) \tilde{\nabla}\cdot \tilde{\textbf{v}} \rangle \approx  \langle \tilde{\textbf{v}} \rangle \cdot \tilde{\nabla} g(\tilde{r}) + g(\tilde{r}) \langle \tilde{\nabla} \cdot \textbf{v} \rangle .
\end{equation}
A general flow field can be separated into compressional (downstream) and extensional (upstream) regions: in the former regions the particles are pushed toward each other by the flow, so the relative velocity between the two particles is negative; instead, in the extensional sectors, the particles move away from each other, leading to a positive radial velocity.
Within this methodology, the actual relative velocity and the flow field divergence are replaced with their angular-averaged values within compressional and extensional regions.
The angular average is necessary to reduce the original PDE (which is soluble only numerically,and even then poses some computational challenges) to an ODE which is analytically soluble. The price to pay for having analytical solutions is that it is not possible to produce deformed contour plots to highlight the angle dependent pcf.\\

\subsection{Averaged velocities}
Now, we will consider two coefficients which are the result of the average procedure,: $\alpha_c$ for the compressional (downstream) and $\alpha_e$ for the extensional (upstream) zone, which are explicitly introduced and defined in Appendix A. The two coefficients define the influence of the angular coordinates on the radial relative velocity and the flow field divergence as shown in Eq.(\ref{relative_velocity_and_divergence}):
\begin{equation}
\begin{cases}
\langle \tilde{\textbf{v}} \rangle_\text{i} = \alpha_\text{i} (1-A(\tilde{r})) \tilde{r}, \\
\langle \tilde{\nabla} \cdot \textbf{v} \rangle_\text{i} = \alpha_\text{i} \biggl( 3(B(\tilde{r})-A(\tilde{r})) - \tilde{r} \dfrac{d A(\tilde{r})}{d \tilde{r}}\biggr)
\end{cases}
\label{relative_velocity_and_divergence}
\end{equation}
with $\text{i}=c$ for the compressional sector and $\text{i}=e$ for the extensional sector. The corresponding values of $\alpha_c$ and $\alpha_e$ are derived in Appendix A. 
\subsection{Lubrication forces}
The difference between compressional and extensional quadrants is also reflected in the modelling of the lubrication forces through the fitting function $G(\tilde{r})$.
If the particles are getting closer to each other as in the compression sectors, then the squeezing of the liquid between them creates a force which opposes the mutual approach~\cite{Batchelor1976}.
In this case we model $G(r)$ through a polynomial ~\cite{Honig,ZacconeNess} which is a polynomial fit to the rigorous solution to the Stokes equation for the specific case of two particles approaching each other \cite{Brenner}:
\begin{equation}
G_c(\tilde{r}) = \dfrac{6 h^2 + 4h}{6 h^2 + 13 h^2 + 2};
\label{Lubrication_Forces_Compressing}
\end{equation}
where $h = \tilde{r} - 1$ is the surface to surface distance between the particles.\\
It is necessary to recall that the proposed function is valid for the scenario where the particles are approaching. On the other hand, if they are moving away from each other, lubrication forces assume a different form that we could not find across the literature so, for simplicity, we decided to neglect them by imposing
\begin{equation}
G_e(\tilde{r}) = 1.
\label{Lubrication_Forces_Extensional}
\end{equation}
\subsection{Final formulation}
With the above specifications, we arrive at the following form for the dimensionless Smoluchowski equation:
\begin{multline}
\epsilon \biggl[ \dfrac{1}{\tilde{r}^2} \dfrac{\text{d}}{\text{d} \tilde{r}} \biggl( \tilde{r}^2  G_\text{i}(\tilde{r}) \dfrac{\text{d} g_\text{i}(\tilde{r})}{\text{d} \tilde{r}} \biggr) + \dfrac{1}{\tilde{r}^2} \dfrac{\text{d}}{\text{d} \tilde{r}} \biggl( \tilde{r}^2 G_\text{i}(\tilde{r}) \dfrac{\text{d} \tilde{U}}{\text{d} \tilde{r}} \biggr) g_\text{i}(\tilde{r}) + \\ +  G_\text{i}(\tilde{r}) \dfrac{\text{d} \tilde{U}}{\text{d} \tilde{r}} \dfrac{\text{d} g_\text{i}(\tilde{r})}{\text{d} \tilde{r}} \biggr] - 2 \biggl( \langle \tilde{\textbf{v}} \rangle_\text{i} \dfrac{\text{d} g_\text{i}(\tilde{r})}{\text{d}\tilde{r}} + g_\text{i}(\tilde{r}) \langle \tilde{\nabla} \cdot \textbf{v} \rangle_\text{i} \biggr) = 0.
\end{multline}
Finally, we put the equation in the following final form which is the most convenient for the perturbative treatment:
\begin{multline}
\epsilon \biggl[ G_\text{i}(\tilde{r}) \biggl( \dfrac{\text{d}^2 g_\text{i}}{\text{d} \tilde{r}^2}  + \dfrac{2}{\tilde{r}} \dfrac{\text{d} g_\text{i}}{\text{d}\tilde{r}}\biggr) + \dfrac{\text{d} G_\text{i}}{\text{d}\tilde{r}} \dfrac{\text{d} g_\text{i}}{\text{d}\tilde{r}}+ g_\text{i} \dfrac{\text{d} \tilde{U}}{\text{d} \tilde{r}} \dfrac{\text{d} G_\text{i}}{\text{d} \tilde{r}}+ \\ + G_\text{i} \dfrac{\text{d} \tilde{U}}{\text{d} \tilde{r}} \dfrac{\text{d} g_\text{i}}{\text{d} \tilde{r}} + G_\text{i} \biggl( \dfrac{2}{\tilde{r}} \dfrac{\text{d}\tilde{U}}{\text{d} \tilde{r}} + \dfrac{\text{d}^2 \tilde{U}}{\text{d} \tilde{r}^2} \biggr)g_\text{i}(\tilde{r}) \biggr] + \\ -  2 \biggl( \langle \tilde{\textbf{v}} \rangle_\text{i} \dfrac{\text{d} g_\text{i}}{\text{d}\tilde{r}} + g_\text{i} \langle \tilde{\nabla} \cdot \textbf{v} \rangle_\text{i} \biggr) = 0.
\label{Smoluchowski_epsilon_final}
\end{multline}
We briefly recall that, since $\tilde{\textbf{v}}$ is influenced not only by the motion of the sheared fluid, but also by the hydrodynamic disturbance between the particles modelled through functions A and B, its divergence assumes not null values even if the fluid is incompressible, as it can be seen in Eq.(\ref{relative_velocity_and_divergence}).
Since Eq.(\ref{Smoluchowski_epsilon_final}) is a second order differential equation, we need two boundary conditions (BCs).
The first one is the usual far-field BC:
\begin{equation}
g_\text{i}(\tilde{r} \to \infty) = 1.
\label{BC_1}
\end{equation}
The second BC constrains the radial flux to be null when the two particles are in direct contact:
\begin{equation}
G_\text{i}(\tilde{r}_c) \biggl( \dfrac{\text{d}g}{\text{d}\tilde{r}} \biggr)(\tilde{r}_c) + \biggl( G_\text{i}(\tilde{r}_c) \dfrac{\text{d} \tilde{U}}{\text{d}\tilde{r}} - 2 \text{Pe} \langle \tilde{\textbf{v}} \rangle_\text{i} \biggr) g_\text{i}(\tilde{r}_c) = 0,
\label{BC_2}
\end{equation}
where $\tilde{r}_c$ is a value of radial distance sufficiently close to the reference particle; in our calculations we take $\tilde{r}_c = 1 + 5 \times 10^{-5}$.\\

From inspection of Eq.(\ref{Smoluchowski_epsilon_final}) it can immediately be seen that the perturbation parameter is linked to the highest order derivative of the ordinary differential equation (ODE). This means that we are dealing with a \textit{singular perturbation} problem that must be solved by the application of  boundary layer theory \cite{BenderOrszag,Hinch,VanDyke}.\\

The approach consists of the evaluation of two different power series related to two different regions of the radial coordinate domain: the outer layer (in this case farther away from the reference particle), where the solution is slowly changing with $\tilde{r}$, and the inner region (closer to the reference particle), usually called boundary layer, where the solution is steeply and very rapidly changing with the radial coordinate~\cite{BenderOrszag}.
\section{Solution method}
The solution to the boundary-layer problem is a combination of two power series in $\epsilon$. The first one is named \textit{outer solution} and provides the approximate form of the pcf in the outer layer where the solution varies slowly with $\tilde{r}$:  
\begin{equation}
g_\text{i}^{\text{out}}(\tilde{r}) = g_{0,\text{i}}^{\text{out}}(\tilde{r}) + \epsilon  g_{1,\text{i}}^{\text{out}}(\tilde{r}),
\label{outersolution}
\end{equation}
where $g_{0,\text{i}}^{\text{out}}$ is the leading order term, while $g_{1,\text{i}}^{\text{out}}$ is the first order term.\\
The second power series is called \textit{inner solution} and provides the solution in the inner layer of the domain where the solution varies dramatically with respect to variations in $\tilde{r}$.
The first step towards building the power series in the inner layer is the application of a change of variable, called inner transformation, to Eq.(\ref{Smoluchowski_epsilon_final}): 
\begin{equation}
\xi = \dfrac{\tilde{r}-\tilde{r}_c}{\delta(\epsilon)}
\label{innervariable}
\end{equation}
where $\delta(\epsilon)$ is the order of magnitude of the width of the inner layer, the small section where the solution varies quickly.\\

Formally, it is necessary to write the inner solution as a power series in $\delta(\epsilon)$. Using the method of dominant balancing as proposed in \cite{Banetta}, it has been shown that $\delta(\epsilon)\sim \epsilon$. Hence, it is possible to write the power series in the inner layer as:
\begin{equation}
g_\text{i}^{\text{in}} (\xi) = g_{0,\text{i}}^{\text{in}}(\xi) + \epsilon g_{1,\text{i}}^{\text{in}}(\xi).
\end{equation} 
\subsection{Solution evaluation}
Following the same steps reported in \cite{Banetta} we arrive at the following forms for the zero-th and first order terms in the outer layer:
\begin{equation}
g_{0,\text{i}}^{\text{out}} = \dfrac{1}{1-A(\tilde{r})} \exp\biggl[{\int_{\tilde{r}}^{\infty}\biggl(\dfrac{3(B-A)}{\tilde{r}(1-A)}\biggr)}d \tilde{r} \biggr],
\label{leading_order_outer_solution_dusty_plasmas}
\end{equation}
\begin{multline}
g_{1,\text{i}}^{\text{out}} = - g_{0,\text{i}}^{\text{out}} \int_{\tilde{r}}^{\infty} \dfrac{1}{2 \langle \tilde{\textbf{v}} \rangle_\text{i}} \biggl \{ G_\text{i} \biggl[ Y^2 + \dfrac{\text{d} Y}{\text{d} \tilde{r}} + \biggl( \dfrac{2}{\tilde{r}} + \dfrac{\text{d} \tilde{U}}{\text{d} \tilde{r}}\biggr) Y(\tilde{r}) + \\ + \dfrac{\text{d}^2\tilde{U}}{\text{d} \tilde{r}^2} + \dfrac{2}{\tilde{r}}\dfrac{\text{d} \tilde{U}}{\text{d} \tilde{r}} \biggr] + \dfrac{\text{d}G_\text{i}}{\text{d}\tilde{r}}\biggl( Y + \dfrac{\text{d} \tilde{U}}{\text{d}\tilde{r}} \biggr) \biggr\} \text{d} \tilde{r}, 
\label{first_order_outer_solution_dusty_plasmas}
\end{multline}
where $Y = - \langle \tilde{\nabla} \cdot \textbf{v} \rangle_\text{i} / \langle \tilde{\textbf{v}} \rangle_\text{i}$.\\
Following again the same steps reported in~\cite{Banetta} for the zero-th and first order terms in the inner layer, we find the following expressions:
\begin{multline}
g_{0,\text{i}}^{\text{in}} = C_1 + C_0 \int_0^{\xi} \exp \biggl[ \biggl(\int_0^{\xi} 2 \dfrac{\langle \tilde{\textbf{v}}(\epsilon = 0) \rangle_\text{i}}{G(\epsilon = 0)} d\xi' \biggr) \biggr] d\xi'
\label{leading_order_inner_solution_dusty_plasmas}
\end{multline}
\begin{multline}
g_{1,\text{i}}^{\text{in}}(\xi) = C_3 + \int_0^{\xi} \biggl \{ C_2 - \int_0^{\xi} \biggl[ \biggl( \dfrac{2}{(\xi' \epsilon + \tilde{r}_c)} + W(\xi') + \dfrac{G_{r,\text{i}}}{G}\biggr) \times \\ \times \dfrac{\text{d} g_{0,\text{i}}^{\text{in}}}{\text{d} \xi'} - 2 \dfrac{\langle \tilde{\nabla}_{\xi'} \cdot \tilde{\textbf{v}}(\xi') \rangle_\text{i}}{G(\xi')} g_{0,\text{i}}^{\text{in}}(\xi') \biggr] \exp \biggl(\int_0^{\xi} - 2 \dfrac{\langle \tilde{\textbf{v}}(\xi')\rangle_\text{i}}{G(\xi')} d\xi' \biggr) \times \\ \times d\xi' \biggr \} \exp \biggl(\int_0^{\xi} 2 \dfrac{\langle \tilde{\textbf{v}}(\xi')\rangle_\text{i}}{G(\xi')} d\xi' \biggr) d\xi',
\label{first_order_inner_solution_dusty_plasmas}
\end{multline}
where $W = (d\tilde{U}/d\xi)/\delta$ and $G_{r,\text{i}} = \delta^{-1}(dG_\text{i}/d\xi)$.
\subsection{Integration constants evaluation}
\begin{figure}
	\centering
	\includegraphics[width = 1 \linewidth]{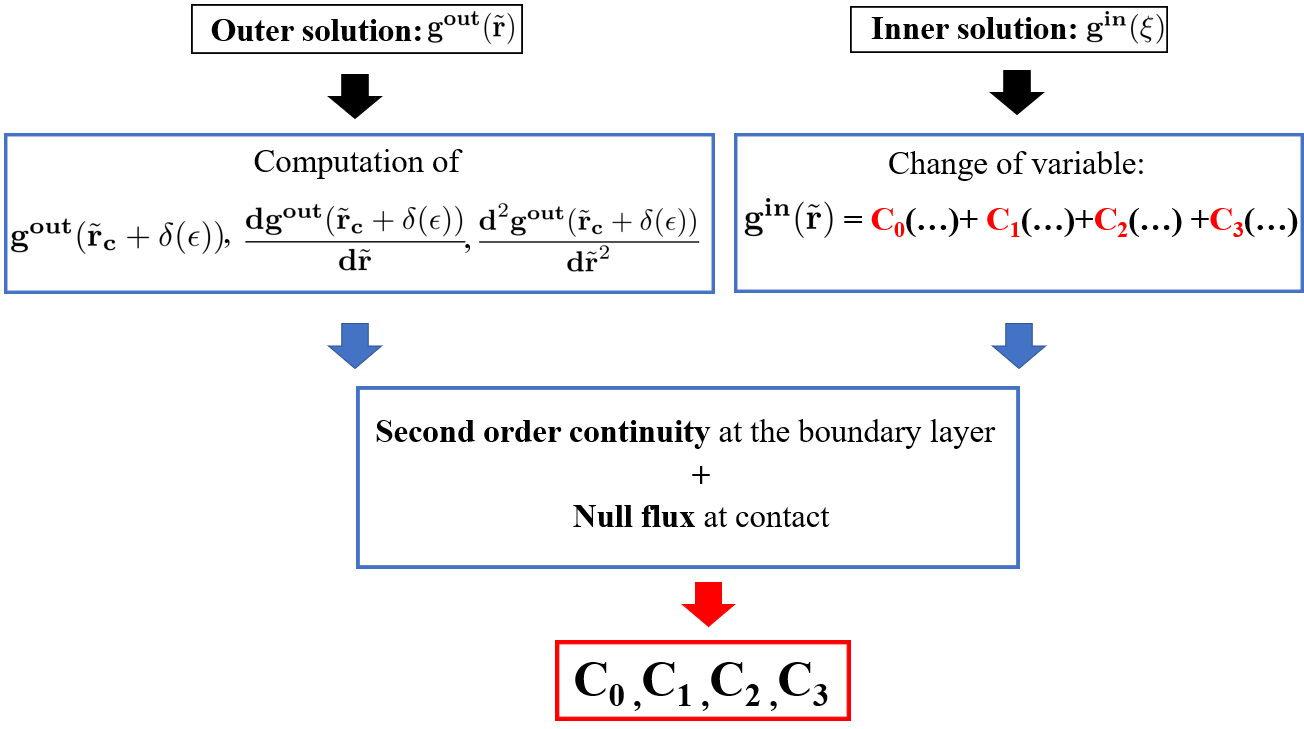}
	\caption{Block diagram with the fundamental steps for the evaluation of the integration constants of $g^{\text{in}}(\tilde{r})$}
	\label{fig:integration_constants_evaluation}
\end{figure}
To summarize, we have evaluated two different power series $g^{\text{in}}$ and $g^{\text{out}}$ which describe the behaviour of the solution in two different adjacent sections of the integration domain, the inner (or boundary) layer and the outer layer, respectively.
The final step to obtain the analytical solution of Eq.(\ref{Smoluchowski_epsilon_final}) is the evaluation of the integration constants $C_0$, $C_1$, $C_2$ and $C_3$ present in the inner solution; the full procedure is summarized in Fig.\ref{fig:integration_constants_evaluation}.\\

Since we have four unknown parameters we need four equations to determine them: the first one will be the condition of zero flux at the reference particle surface, Eq.(\ref{BC_2}), while the other three can be obtained from the so-called $\textit{patching}$ procedure \cite{BenderOrszag}. 
The general principle is as follows. We start from two solutions which share a common border: if one of the two is known and the other has $n$ constants to be evaluated, it is necessary to apply a condition of continuity of order $n-1$.\\

This principle is suitable for our case since we know the full behaviour of the outer solution and we have three remaining conditions to be fixed in order to find the three remaining constants. Hence, we need to fix a second order continuity condition between $g^{\text{out}}$ and $g^{\text{in}}$ at their shared border, that is 
$\tilde{r} = \tilde{r}_c + \epsilon$.
After having obtained the complete structure of the inner solution, we need to group together all the terms which multiply the same integration constant.\\

Finally, it is possible to evaluate all the integration constants by solving the following linear system which arises from the patching procedure,
\begin{equation}
\begin{cases}
g_\text{i}^{\text{out}}(\tilde{r} = \tilde{r}_c + \epsilon) = g_\text{i}^{\text{in}}(\tilde{r} = \tilde{r}_c + \epsilon) \\
\dfrac{\text{d} g_\text{i}^{\text{out}}(\tilde{r} = \tilde{r}_c + \epsilon)}{\text{d} \tilde{r}} = \dfrac{\text{d} g_\text{i}^{\text{in}}(\tilde{r} = \tilde{r}_c + \epsilon)}{\text{d} \tilde{r}} \\
\dfrac{\text{d}^2 g_\text{i}^{\text{out}}(\tilde{r} = \tilde{r}_c + \epsilon)}{\text{d} \tilde{r}^2} = \dfrac{\text{d}^2 g_\text{i}^{\text{in}}(\tilde{r} = \tilde{r}_c + \epsilon)}{\text{d} \tilde{r}^2}
\end{cases}
\label{Integration_Constants_Evaluation}
\end{equation}
together with the application of Eq.(\ref{BC_2}); from the solution of this linear system we evaluate the four integration constants $C_0$, $C_1$, $C_2$ and $C_3$ as functions of the P\'eclet number which will lead to the final form of $g^{\text{in}}$.\\
\section{Results}
\begin{figure*}
	\centering
	\subfloat{\includegraphics[width = 0.47 \linewidth]{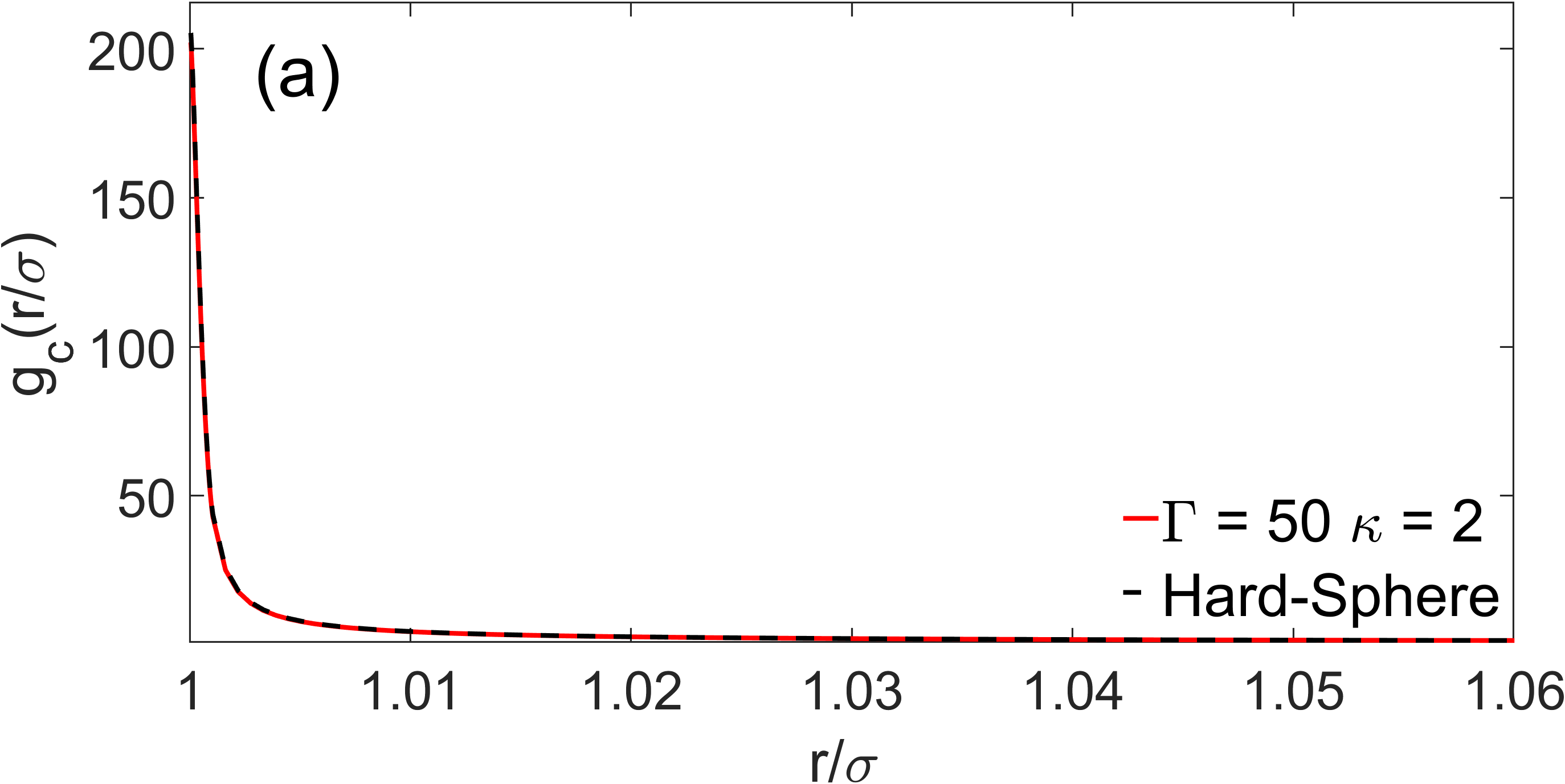}} \quad 
	\subfloat{\includegraphics[width = 0.462 \linewidth]{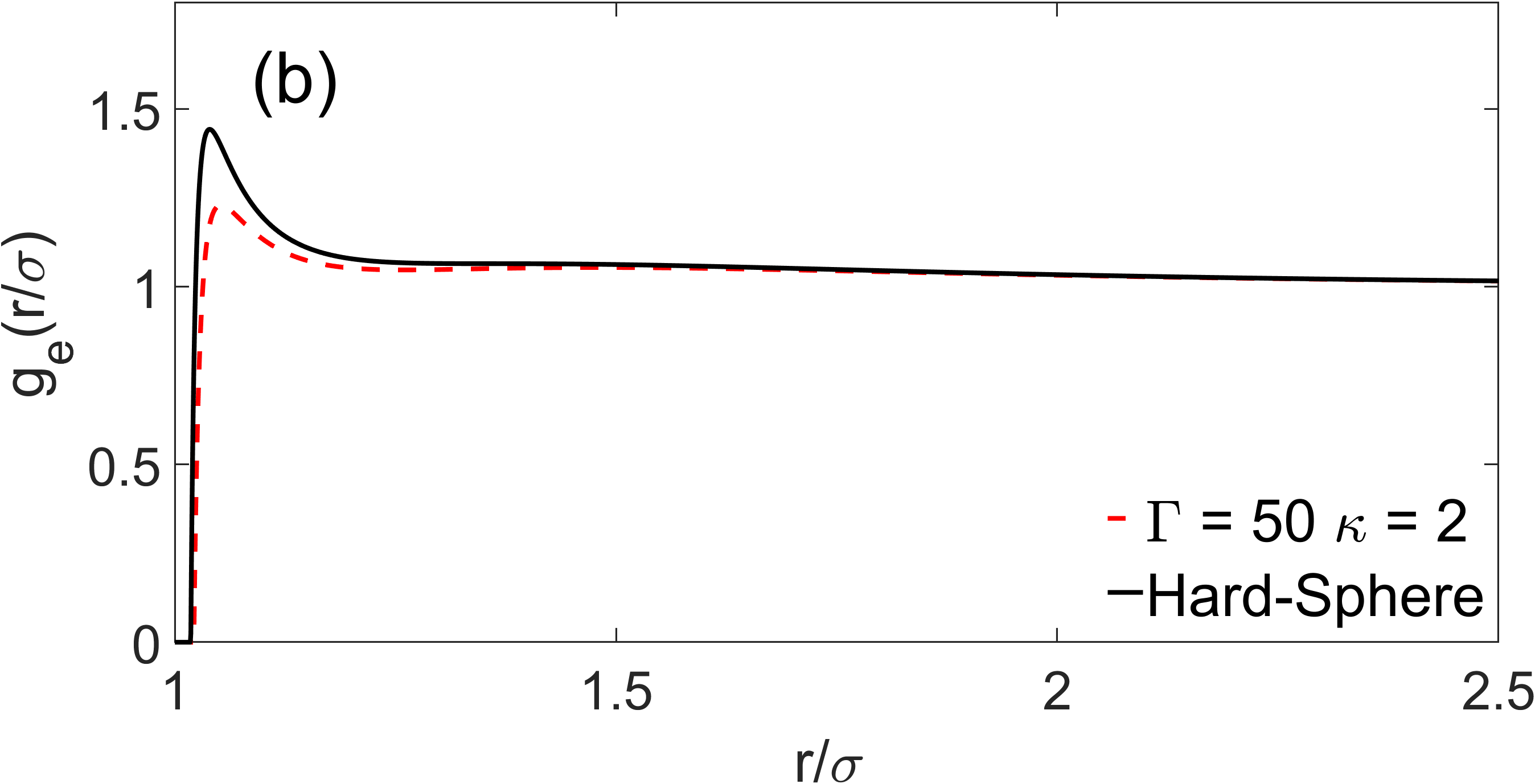}}
	\caption{Effect of the repulsive DH potential on the pair correlation function in the compressing $g_c(\tilde{r})$ and extensional quadrants $g_e(\tilde{r})$ of a strongly sheared suspension (Pe = 1000).}
	\label{fig:High_Peclet}
\end{figure*}
Since we are interested in studying the micro-structure of charge-stabilized particles, we implement the screened-Coulomb Debye-H\"uckel (or Yukawa) interaction potential with the addition of an hard-sphere wall which is epxressed by infinte values of $\tilde{U}(\tilde{r})$ if $\tilde{r} < \tilde{r}_c$:
\begin{equation}
\begin{cases}
\tilde{U}(\tilde{r}) = \infty \quad \tilde{r} < \tilde{r}_c;\\
\tilde{U}(\tilde{r}) = \dfrac{(Z^*e)^2}{4 \pi \epsilon_{r} \epsilon_{0} \sigma  k_B T} \dfrac{\exp( - \kappa \tilde{r})}{\tilde{r}} = \Gamma \dfrac{\exp( - \kappa \tilde{r})}{\tilde{r}} \quad \tilde{r} > \tilde{r}_c,
\end{cases}
\label{Yukawa_Potential}
\end{equation}
where $Z^*$ is the effective charge, $e$ the electron charge, $\epsilon_{r}$ the medium relative dielectric permittivity, $\epsilon_{0}$ the dielectric permittivity of vacuum, and  $\kappa$  the dimensionless Debye screening parameter in units of $\sigma^{-1}$. The inverse of $\kappa$ is the Debye length $\kappa^{-1}$ (with units of $\sigma$), which is the length scale within which the interactions are non-negligible. In a colloidal suspension, $\kappa$  is a function of the ionic strength. The parameter $\Gamma$ is known as the coupling constant and controls the strength of the (screened) Coulomb repulsion.\\

The analytical approach has been validated in a parameter-free comparison with numerical simulations data of hard spheres in\cite{Banetta}.\\

In the next section we present predictions of the pcf in the compressing and the extensional sectors at different values of P\'eclet numbers (in the regime $Pe \gg 1$) and upon varying the control parameters of the DH potential (i.e. $\kappa$ and $\Gamma$).

\subsection{\textbf{High P\'eclet numbers}}
In Fig.\ref{fig:High_Peclet} we present the locally averaged pcf for both compression and extensional sectors at $Pe = 1000$. In this case the interactions play a completely negligible role in both the compressing and extensional regions where only the interplay between the flow field and lubrication forces determines the pcf. In this limit, the solution is strongly dominated by the outer layer, which, in turn, is totally dominated by the hydrodynamics.\\

In Fig.\ref{fig:High_Peclet}a the pcf in the compression sectors is shown. We observe a two orders of magnitude increase of the pcf near the surface of the reference particle, because the strong compressing effect of the flow field in these regions  pushes the particles towards each other.
On the other hand, in the extensional sectors, as shown in Fig. Fig.\ref{fig:High_Peclet}b, the flow field tends to flatten the pcf out to unity (homogeneous concentration): this makes sense because the particles are not influenced by the inter-particle interactions, since the shear induced effects are three orders of magnitude more dominant than the Brownian-induced ones; nor are they influenced by lubrication forces since these take place when the fluid between the particles is squeezed, which happens in the compression quadrants only.
Only a comparatively much smaller maximum is seen in the extensional sectors, which is due to the competition between the action of the flow, which tends to push particles away from each other, and the effect of the hydrodynamic disturbances due to the relative motion of the particles, as encoded in the hydrodynamic functions $A(\tilde{r})$ and $B(\tilde{r})$.
\\

\subsection{\textbf{Intermediate P\'eclet numbers (Pe=10)}}
For colloidal suspensions experiencing weaker shear rates, the interaction potential plays a non-negligible role in determining the microstructure in both the compressing and extensional quadrants. Furthermore, the interplay between interaction potential, flow field and hydrodynamic(lubrication) interactions give rise to new phenomena.\\
\subsubsection{Compressing quadrants}
\begin{figure*}
	\centering
	\subfloat{\includegraphics[width = 0.47 \linewidth]{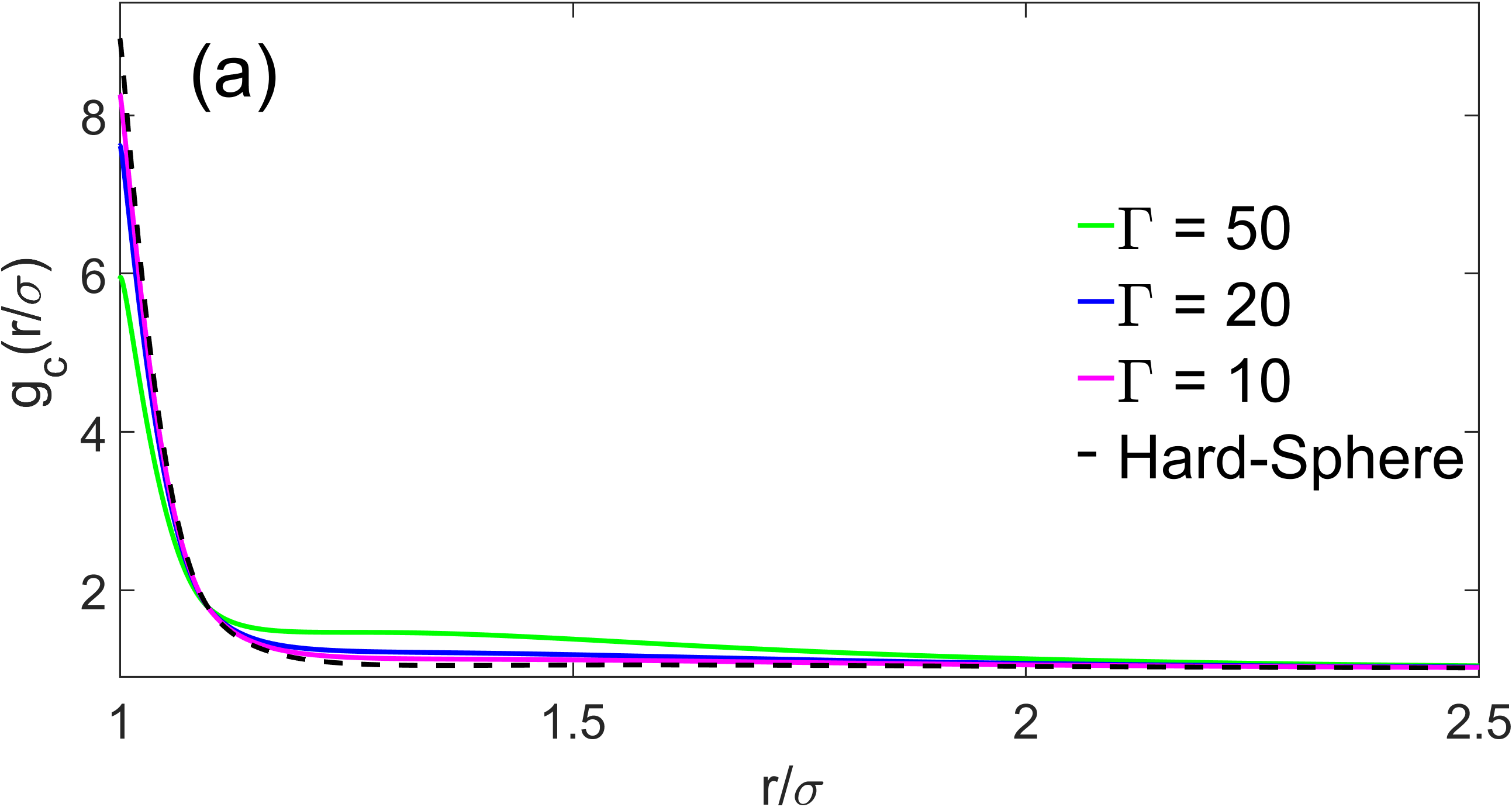}} \quad 
	\subfloat{\includegraphics[width = 0.475 \linewidth]{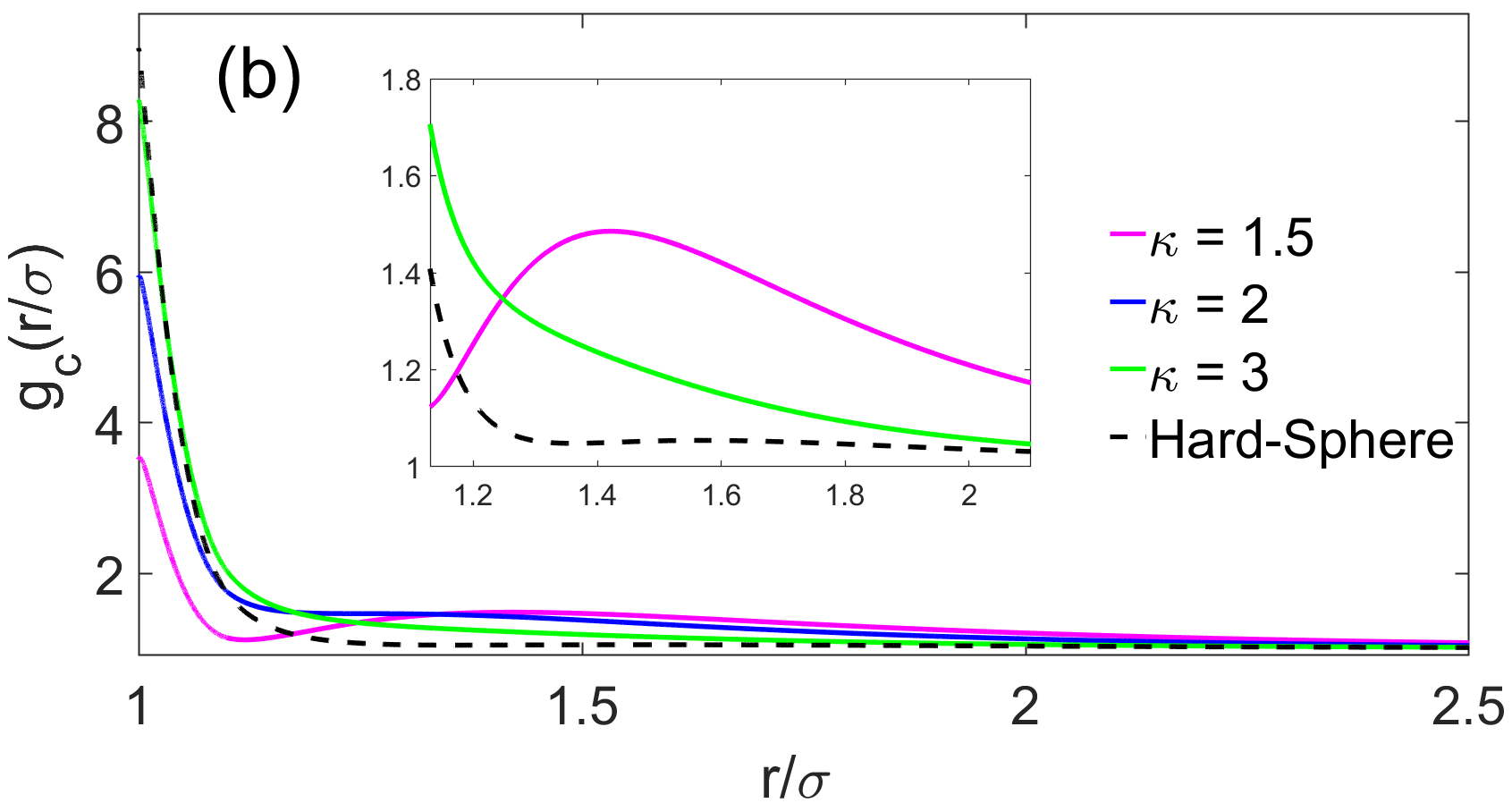}}
	\caption{Effect of the repulsive DH potential on $g_c(\tilde{r})$ at low values of the P\'eclet number (Pe = 10): a) Effect of varying the coupling parameter $\Gamma$ with $\kappa = 2$ fixed, b) Effect of varying  the Debye parameter with $\Gamma = 50$ kept fixed.}
	\label{fig:Low_Pe_Compressing}
\end{figure*}
Results for the pcf in compression sectors are reported in Fig. 4. In the compression sectors, where the particles are pushed towards each other by the flow, we observe an accumulation peak with values of $g_c(\tilde{r})$ bigger than unity near contact, which means that we have an increased probability of finding particles near the reference one.\\

In Fig. 4(a) the contribution of the repulsive DH interaction causes the peak of the pcf to decrease with the increase of the coupling parameter $\Gamma$, which controls the repulsion strength. Clearly, the screened-Coulomb DH repulsion opposes a resistance to the approach of the particles and it has been shown  in  \cite{NazockdastMorrisSoftSpheres} that the conservative interactions play a dominant role, also over the lubrication forces, in the determination of the peak at contact.\\

As a consistency check, we see that as $\Gamma$ goes to zero or $\kappa$ goes to infinity, the microstructure of repulsive interacting particles gets closer to the hard-sphere limit, a clear evidence of the good reliability of the presented method.\\

In Fig. 4(b) we present results for the pcf in the compressional sectors, this time upon varying the Debye screening parameter $\kappa$. Increasing $\kappa$ means decreasing the Debye screening length $\kappa^{-1}$, which sets the length scale for the decay of the DH repulsion. A new effect is predicted here for the first time: as the Debye length decreases, a secondary maximum appears for $\kappa=1.5$ (in units of $\sigma$) at a position $r=1.4 \sigma$, which is slightly less than the Debye length $(1+\kappa^{-1})\sigma \approx 1.67 \sigma$. This effect can be interpreted as a local "accumulation" of particles advected by the flow towards the electrostatic repulsive wall. If the Debye length is too short compared to the primary accumulation peak, this effect cannot be seen. 
\subsubsection{Extensional quadrants}
\begin{figure*}
	\centering
	\subfloat{\includegraphics[width = 0.47 \linewidth]{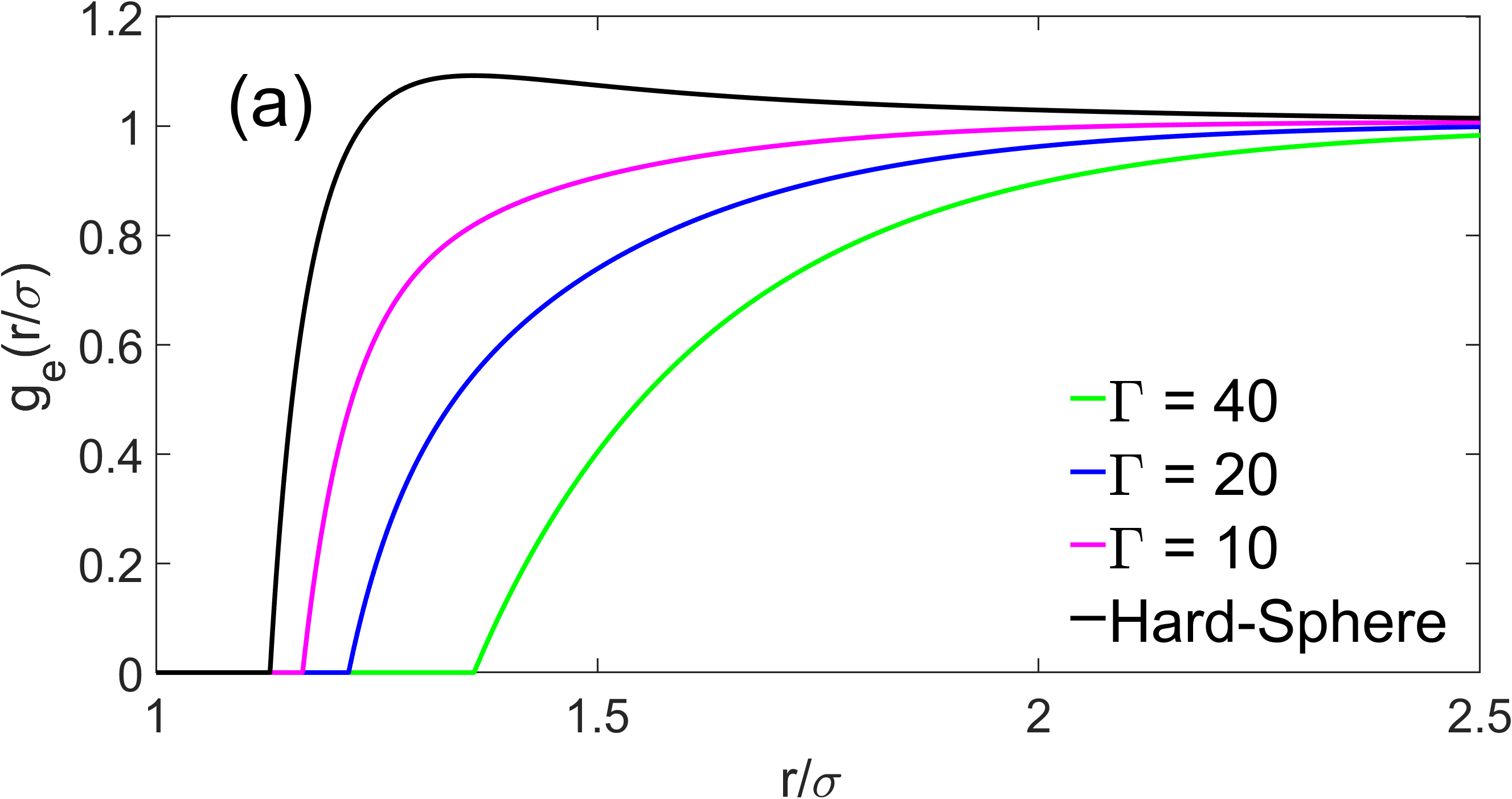}} \quad
	\subfloat{\includegraphics[width = 0.47 \linewidth]{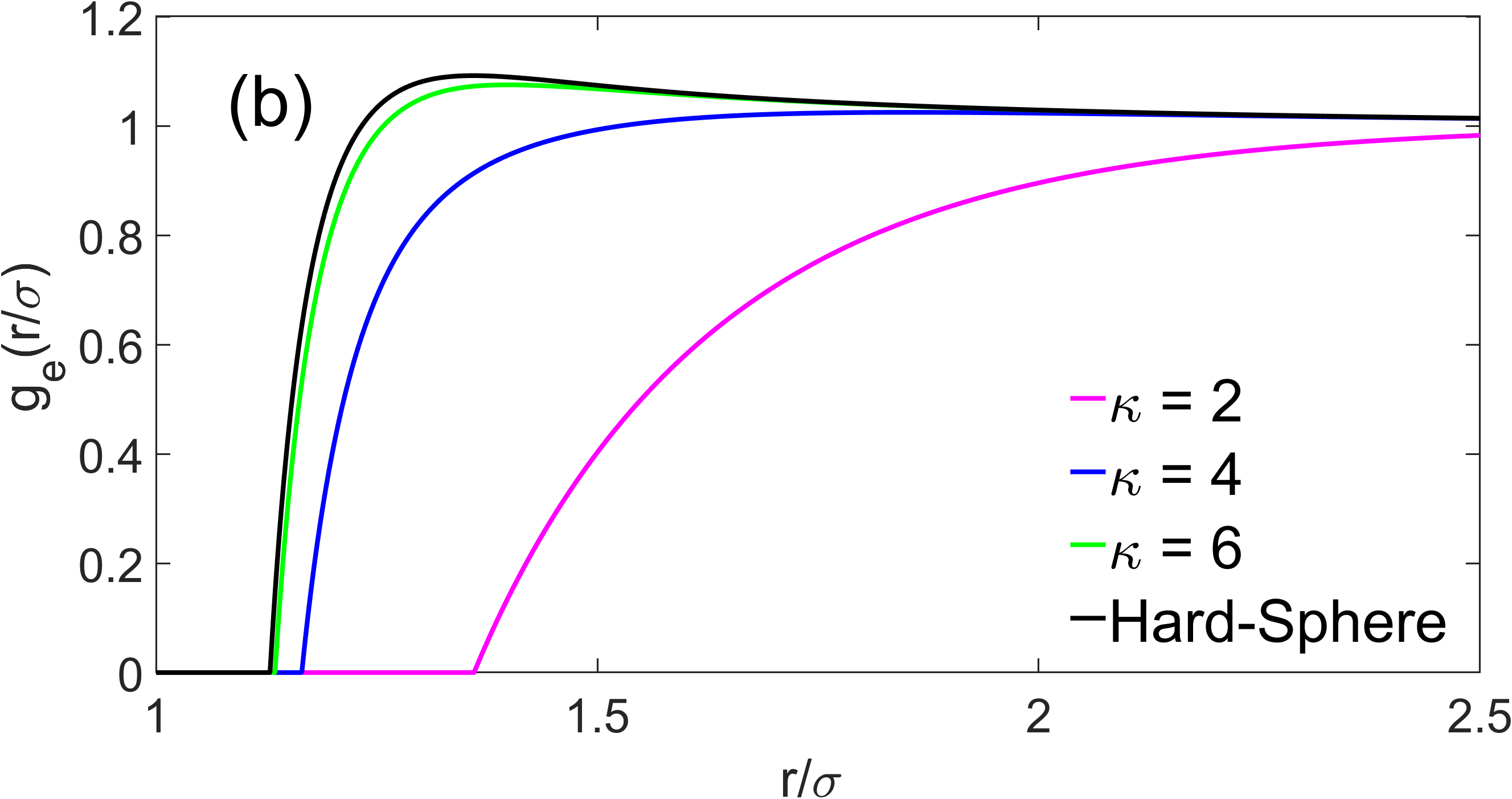}}
	\caption{Effect of the repulsive DH interaction on $g_e(\tilde{r})$ at low values of the P\'eclet number (Pe = 10): a) Effect of varying the coupling parameter $\Gamma$ with $\kappa = 2$ fixed, b) Effect of varying the Debye parameter $\kappa$ with $\Gamma = 40$ kept fixed.}
	\label{fig:Low_Pe_extensional}
\end{figure*}
Now we focus our attention on Fig.\ref{fig:Low_Pe_extensional} where we present the pcf in the extensional quadrants. First of all, we notice that there is no peak or increased probability of finding particles near the surface of the reference particle, which has been seen also for hard spheres\cite{MorrisBrady}. This result is physically meaningful since the flow field causes the particles to move away from each other.\\

On the other hand, and this is a new effect seen here for the first time, the synergy between the (locally extensional, ''repulsive-like") flow and the screened-Coulomb interaction leads to a \textit{depletion layer} near contact, within which the probability of finding a particle is identically zero. The width of the depletion layer increases with the increase of either the strength of the repulsion, controlled by $\Gamma$, or the range of the repulsion, controlled by $\kappa^{-1}$.
Also in this case, if we decrease the Debye length $\kappa^{-1}$, or decrease $\Gamma$, the micro-structure tends to approach the behaviour of hard-spheres.

\section{Conclusion and future steps}
In this work we presented an analytical theory of the pair correlation function of charge-stabilized colloids in shear flow, using the intermediate-aymptotics solution method to the Smoluchowski equation that was recently developed in~\cite{Banetta}. The theory has been built on a series of hypothesis:
\begin{enumerate}
	\item The tangential contribution of the lubrication forces with respect to the line of centres acting on Brownian motion has been neglected;
	
	\item Integral average over two different domains of the solid angle: compressing quadrants, where the particles approach each other, and extensional regions, where they fade away from each other;
	
	\item Decoupling approximation: the average of the scalar product is sufficiently close to the scalar product of the averages.
\end{enumerate}
The method yields the locally averaged pair correlation function for the compression and extensional sectors of the solid angle. In the compression sectors, an accumulation peak near contact is visible, which can be lowered upon increasing the repulsion parameters of the Debye-H\"uckel  potential. As the Debye length increases (and becomes larger than the particle diameter $\sigma$) a secondary maximum appears which is due to the competition between the advecting action of the flow (pushing particles against each other, hence attractive-like) and the effect of screened electrostatic repulsion. The secondary maximum occurs at separations comparable to the Debye length. 
In the extensional sectors, instead, no accumulation peak is visible, due to the action of the flow that tends to move particles away from each other in the extensional sectors. Instead, the occurrence of a depletion layer, where the pair correlation function is identically zero, is predicted. The width of the depletion layer increases upon increasing either the charge repulsion strength or the Debye length. 

In future work some predictions from the theory, such as the presence of a depletion layer in extensional quadrants, can be confirmed experimentally by evaluating the structure factor, and the pcf through the appropriate Fourier-transform, of known repulsive systems under consistent sheared conditions. Moreover, this methodology will be applied to low-Peclet number conditions, where the inter-particle interactions play the most dominant role. Across the literature there is already experimental evidence related to the spatial arrangement of this type of systems under weakly sheared conditions which can be utilized as a possible validation \cite{Nosenko}. Furthermore, this analytical theory can serve as the starting point for predictions of viscosity and rheology of sheared colloidal suspensions, as well as in systems such as plasmas and dusty plasmas~\cite{Murillo}. In the case of colloidal systems, this approach could be combined, in future work, with Mode-Coupling Theory~\cite{Ballauff1,Naegele,Ballauff2,Ballauff3,Schall} to arrive at predictions of dynamics and rheological response of interacting colloidal particles under strong shear flows; also, it could be used to predict and model controlled self-assembly of nanoparticles using shear flow~\cite{Likos}.

\section{Acknowledgements}
This work is dedicated to Prof. Matthias Ballauff in occasion of his retirement. Prof. Massimo Morbidelli is gratefully acknowledged for many inspiring discussions and for providing motivation to study this problem. L.B. gratefully acknowledges financial support from Synthomer UK Ltd.
\appendix
\section{Mathematical formalism}
Let's focus or attention on  the Brownian contribution to Eq.(\ref{Smoluchowski_epsilon_1}):
\begin{multline}
\tilde{\nabla} \cdot \biggl( \underline{\tilde{\textbf{D}}}^\text{Br} \cdot \tilde{\nabla} g(\tilde{\textbf{r}}) \biggr) - \tilde{\nabla} \cdot  \biggl[ \biggl( \underline{\tilde{\textbf{D}}}^\text{Br} \cdot   \tilde{\textbf{K}}^\text{int} \biggr) g(\tilde{\textbf{r}}) \biggr] = \\ 
= \tilde{\nabla} \cdot \biggl( \underline{\tilde{\textbf{D}}}^\text{Br} \cdot \tilde{\nabla} g(\tilde{\textbf{r}}) \biggr) - \tilde{\nabla} \cdot  \biggl( \underline{\tilde{\textbf{D}}}^\text{Br} \cdot   \tilde{\textbf{K}}^\text{int} \biggr) g(\tilde{\textbf{r}}) - \\ \biggl( \underline{\tilde{\textbf{D}}}^\text{Br} \cdot   \tilde{\textbf{K}}^\text{int} \biggr) \cdot \tilde{\nabla} g(\tilde{\textbf{r}})
\label{Brownian_Contribution_1}
\end{multline}
Expressing all the components and the divergence operator we obtain, respectively
\begin{multline}
\tilde{\nabla} \cdot \begin{bmatrix} G(r) \dfrac{\partial g(\tilde{\textbf{r}})}{\partial \tilde{r}} \\ \\ \dfrac{H(r)}{\tilde{r}} \dfrac{\partial g(\tilde{\textbf{r}})}{\partial \theta} \\ \\ \dfrac{H(\tilde{r})}{\tilde{r} \sin{\theta}} \dfrac{\partial g(\tilde{\textbf{r}})}{\partial \phi} \end{bmatrix} - \tilde{\nabla} \cdot \begin{bmatrix}   - G(\tilde{r}) \dfrac{d \tilde{U}}{d \tilde{r}} \\ \\ 0 \\ \\ 0\end{bmatrix} g(\tilde{\textbf{r}}) + \\ - \begin{bmatrix}   - G(\tilde{r}) \dfrac{d \tilde{U}}{d \tilde{r}} \\ \\ 0 \\ \\ 0\end{bmatrix} \cdot \begin{bmatrix}  \dfrac{\partial g(\textbf{r})}{\partial \tilde{r}} \\ \\ \dfrac{1}{\tilde{r}}\dfrac{\partial g(\textbf{r})}{\partial \theta} \\ \\ \dfrac{1}{\tilde{r} \sin{\theta}}\dfrac{\partial g(\textbf{r})}{\partial \phi} \end{bmatrix}
\label{Brownian_Contribution_1}
\end{multline}
and
\begin{multline}
\dfrac{1}{\tilde{r}^2} \dfrac{\partial}{\partial \tilde{r}} \biggl( \tilde{r}^2 G(\tilde{r}) \dfrac{\partial g(\tilde{\textbf{r}})}{\partial \tilde{r}} \biggr) + \dfrac{H(\tilde{r})}{\tilde{r}^2 \sin{\theta}} \biggl[ \dfrac{\partial}{\partial \theta} \biggl( \sin{\theta} \dfrac{\partial g(\tilde{\textbf{r}})}{\partial \theta} \biggr) + \\ + \dfrac{\partial^2 g(\tilde{\textbf{r}})}{\partial \phi^2}  \biggr] +  \dfrac{1}{\tilde{r}^2} \dfrac{\partial}{\partial \tilde{r}} \biggl( \tilde{r}^2 G(\tilde{r}) \dfrac{\partial \tilde{U}}{\partial \tilde{r}} \biggr) g(\tilde{\textbf{r}}) + G(\tilde{r}) \dfrac{d \tilde{U}}{d \tilde{r}} \dfrac{\partial g(\tilde{\textbf{r}})}{\partial \tilde{r}}
\end{multline}
If we neglect the lubrication forces acting on tangential directions we end up with
\begin{multline}
\dfrac{1}{\tilde{r}^2} \dfrac{\partial}{\partial \tilde{r}} \biggl( \tilde{r}^2 G(\tilde{r}) \dfrac{\partial g(\tilde{\textbf{r}})}{\partial \tilde{r}} \biggr) + \dfrac{1}{\tilde{r}^2} \dfrac{\partial}{\partial \tilde{r}} \biggl( \tilde{r}^2 G(\tilde{r}) \dfrac{d \tilde{U}}{d \tilde{r}} \biggr) g(\tilde{\textbf{r}}) + \\ +G(\tilde{r}) \dfrac{d \tilde{U}}{d \tilde{r}} \dfrac{\partial g(\tilde{\textbf{r}})}{\partial \tilde{r}}.
\end{multline}
Since every contribution from the angular coordinates disappeared it is possible to apply the angular average directly on the pcf on this portion of Eq.(\ref{Smoluchowski_epsilon_2}).
\section{Angular averaging}
In this section we describe the procedure where we describe the angular averaging procedure with which we  evaluate $\langle \tilde{\textbf{v}} \rangle$ and $\langle  \tilde{\nabla} \cdot \tilde{\textbf{v}} \rangle$.
We start the procedure from Eq.(\ref{dimensionless_velocity_axisymmetric})
\begin{equation}
\begin{cases}
\tilde{v}_r = \tilde{r} (1-A(\tilde{r}))\sin^2{\theta} \sin\phi \cos\phi \\
\tilde{v}_{\theta} = \tilde{r} (1- B(\tilde{r})) \sin{\theta} \cos{\theta}\sin\phi \cos\phi,\\
\tilde{v}_{\phi} = \tilde{r} \sin\theta\biggl( \cos^2\phi - \dfrac{B(\tilde{r})}{2} \cos(2 \phi) \biggr)
\end{cases}
\label{dimensionless_velocity_axisymmetric}
\end{equation}
where $A(\tilde{r})$ and $B(\tilde{r})$ are functions representing the effect of the hydrodynamic disturbance along the radial and angular coordinate, respectively. Their values can be taken from the literature \cite{Batchelor} and, in order to use them in the present analytical calculations, they are fitted through the following algebraic expressions \cite{Melis}:
\begin{equation}
\begin{cases}
A(\tilde{r}) = \dfrac{113.2568894}{(2 \tilde{r})^5} +\dfrac{307.8264828}{(2 \tilde{r})^6} +\\
- \dfrac{2607.54064288}{(2 \tilde{r})^7} + \dfrac{3333.72020041}{(2 \tilde{r})^8} \\\\

B(\tilde{r}) = \dfrac{0.96337157}{(2 \tilde{r} - 1.90461683)^{1.99517070}} +\\
- \dfrac{0.93850774}{(2 \tilde{r} - 1.90378420)^{2.01254004}}.
\label{hydrodynamic_functions_fitting}
\end{cases}
\end{equation}
Our goal is to evaluate the average radial velocity in the area where the particles are approaching each other, which means the ensemble of angular coordinates $\tilde{v}_r < 0 $.\\
It is found that the above mentioned condition is satisfied, for $\tilde{r} >0$, $\forall \theta \in [0,\pi]$, $\phi \in [\pi/2,\pi]$ and $\phi \in [3\pi/2, 2\pi]$. Now we apply the angular average obtaining:
\begin{multline}
\langle \tilde{\textbf{v}} \rangle_\text{c} =  \tilde{r} (1-A(\tilde{r}))  \dfrac{1}{4 \pi}  \biggl[\int_0^{\pi} \sin^2(\theta) \sin\theta d\theta \times \\ \times \biggl( \int_{\pi/2}^{\pi} \sin(\phi) \cos(\phi) d \phi + \int_{3\pi/2}^{2 \pi} \sin(\phi) \cos(\phi) d \phi \biggr) \biggr].
\label{angular_average_relative_velocity_compression}
\end{multline}
Through this procedure we can obtain
\begin{equation}
\alpha_\text{c} = -\dfrac{1}{3 \pi}.
\end{equation}
To find the upstream region we need to impose $\tilde{v}_r>0$, which is given by  $\forall \theta \in [0,\pi]$, $\phi \in [0,\pi/2]$ and $\phi \in [\pi, 3\pi/2]$. Applying the same procedure seen before for $\alpha_c$ we obtain:
\begin{multline}
\langle \tilde{\textbf{v}} \rangle_\text{e} =  \tilde{r} (1-A(\tilde{r}))  \dfrac{1}{4 \pi}  \biggl[\int_0^{\pi} \sin^2(\theta) \sin\theta d\theta \times \\ \times \biggl( \int_{0}^{\pi/2} \sin(\phi) \cos(\phi) d \phi + \int_{\pi}^{ 3\pi/2} \sin(\phi) \cos(\phi) d \phi \biggr) \biggr],
\label{angular_average_relative_velocity_extension}
\end{multline}
and, as a consequence
\begin{equation}
\alpha_\text{e} = \dfrac{1}{3 \pi}.
\end{equation}
From this point onward we will consider the compressional case only; the extensional one can be derived in a straightforward manner by replacing $\alpha_c$ with $\alpha_e$.\\
Next we consider the divergence of the flow field, which can be written in spherical coordinates as
\begin{multline}
\tilde{\nabla} \cdot \tilde{\textbf{v}} = \\ =\dfrac{1}{\tilde{r}^2} \dfrac{\partial}{\partial \tilde{r}} \biggl( \tilde{r}^2 \tilde{v}_r \biggr) + \dfrac{1}{\tilde{r} \sin({\theta})} \dfrac{\partial}{\partial \theta} \biggl( \sin{\theta} v_{\theta} \biggr) + \dfrac{1}{\tilde{r} \sin{\theta}} \dfrac{\partial}{\partial \phi}\biggl( v_{\phi} \biggr).
\end{multline}
Adopting the correlations in Eq.(\ref{dimensionless_velocity_axisymmetric}), we can evaluate the divergence as
\begin{equation}
\tilde{\nabla} \cdot \tilde{\textbf{v}} = \biggl[ 3 (B(\tilde{r})-A(\tilde{r}))-\tilde{r} \dfrac{\text{d}A}{\text{d}\tilde{r}} \biggr] \sin^2{\theta} \sin{\phi} \cos{\phi}.
\end{equation}
Finally, we apply the integral average previously seen for $\langle \textbf{v} \rangle_\text{i}$ and we obtain:
\begin{equation}
\langle  \tilde{\nabla} \cdot \tilde{\textbf{v}} \rangle_\text{i} = \alpha_i \biggl[ 3 (B(\tilde{r})-A(\tilde{r}))-\tilde{r} \dfrac{\text{d}A}{\text{d}\tilde{r}} \biggr],
\label{angular_averaged_divergence}
\end{equation}
with $\text{i} = \text{c,e}$ for compression (c) and extension (e), respectively.
\bibliographystyle{plain}

\begin{thebibliography}{99}
	\bibitem{Larsen}
	J. Larsen, \textit{Foundations of High-energy-density Physics: Physical Processes of Matter at Extreme Conditions}, (Cambridge University Press, 2017).
	
	\bibitem{Hansen}
	J.P. Hansen, \textit{Statistical mechanics of dense ionized matter. I. Equilibrium properties of the classical one-component plasma}, Physical Review A, \textbf{8}, 3096-3109 (1973).
	
	\bibitem{Dhont_Book}
	J.K.G. Dhont, \textit{An introduction to dynamics of colloids}, (Elsevier Science,1996).
	
	\bibitem{Fortov}
	V.E. Fortov, O.E. Petrov, O.S. Vaulina, K.G. Koss, \textit{Brownian motion of dust particles in a weakly ionized plasma}, JETP Letters \textbf{97}, 322-326.
	
	\bibitem{Mura}
	F. Mura and A. Zaccone, \textit{Effects of shear flow on phase nucleation and crystallization}, Physical Review E 93, 042803 (2016).
	
	\bibitem{Guido}
	V. Preziosi, A. Perazzo, G. Tomaiuolo, V. Pipich, D. Danino, L. Paduano, S. Guido, \textit{Flow-induced nanostructuring of gelled emulsions}, Soft Matter 13, 5696 (2017).
	
	\bibitem{Lowen}
	H.L$\ddot{o}$wen, J.P. Hansen, J.N. Roux, \textit{Brownian dynamics and kinetic glass transition in colloidal suspoensions}, Physical Review A \textbf{44}, 1169-1181 (1991).
	
	\bibitem{Falkovich}
	G. Falkovich, A. Fouxon, M. G. Stepanov, \textit{Acceleration of rain initiation by cloud turbulence}, Nature 419, 151–154 (2002).
	
	\bibitem{Batchelor}
	G. K. Batchelor, J. T. Green, \textit{The determination of the bulk stress in a suspension of spherical particles to order} $c^2$, Journal of Fluid Mechanics \textbf{56}, 401-427 (1972).
	
	\bibitem{BatchelorGreen}
	G.K. Batchelor, J.T. Green, \textit{The hydrodynamic interaction of two small freely-moving spheres in a linear flow field},Journal of Fluid Mechanics \textbf{56}, 375-400(1972).
	
	\bibitem{Blawdz} 
	J. Blawzdziewicz and G. Szamel, \textit{Structure and rheology of
		semidilute suspension in shear flow}, Phys. Rev. E 48, 4632 (1993).
	
	\bibitem{MorrisBrady}
	J. F. Brady, J. F. Morris, \textit{Microstructure of strongly sheared suspensions and its impact on rheology and diffusion}, Journal of Fluid Mechanics \textbf{348}, 103-139 (1997).
	
	\bibitem{BradyBossis}
	J. F. Brady, G. Bossis \textit{Stokesian dynamics}, Annual Reviews of Fluid Mechanics \textbf{74}, 111 - 157(1988).
	
	\bibitem{Morris}
	J. F. Morris, B. Katyal, \textit{Microstructure from simulated Brownian suspension flows at large shear rate}, Physics of Fluids \textbf{14}, 1920-1937 (2002).
	
	\bibitem{NazockdastMorris}
	E.Nazockdast, J.F. Morris,
	\textit{Microstructural theory and the rheology of concentrated colloidal suspensions},
	Journal of Fluid Mechanics \textbf{713}, 420-452 (2012).
	
	\bibitem{NazockdastMorrisSoftSpheres}
	E.Nazockdast, J.F. Morris,
	\textit{Effect of repulsive interactions on structure and rheology of sheared colloidal dispersions},
	Soft Matter \textbf{8}, 4223-4234 (2012).
	
	\bibitem{Banetta}
	L. Banetta, A.Zaccone, \textit{Radial Distribution Function of Lennard Jones Fluids in shear flows from intermediate asymptotics}, Physical Review E \textbf{99}, 052606 (2019).
	
	\bibitem{Brenner}
	H. Brenner,\textit{The slow motion of  a sphere through a viscous fluid towards a plane surface}, Chemical Engineering Science \textbf{6}, 242-251 (1961).
	
	\bibitem{Honig}
	E. P. Honig, G. J. Roebersen, and P. H. Wieresema, \textit{Effect of hydrodynamic interaction on the coagulation rate of hydrophobic colloids}, J. Coll. Interface Sci. 36, 97 (1971).
	
	\bibitem{Adler}
	P.M. Adler, \textit{Interaction of unequal spheres. I. Hydrodynamic interaction: colloidal forces}, Journal of colloidal and interface science \textbf{84}, 461-473 (1981).
	
	\bibitem{Lin}
	C.J. Lin, K.J. Lee, N.F. Sather, \textit{Slow motion of two spheres in a shear field}, Journal of Fluid Mechanics \textbf{43}, 35-57, (1970).
	
	\bibitem{BenderOrszag}
	C. M. Bender, S.A. Orszag, \textit{Advanced mathematical methods for scientists and engineers I: Asymptotic methods and perturbation theory} (Springer Science \& Business Media, New York, 1999).
	
	\bibitem{Hinch}
	J.Hinch, \textit{Perturbation methods}, (Cambridge University Press, 1991).
	
	\bibitem{VanDyke}
	M. Van Dyke, \textit{Perturbation methods in fluid mechanics}, (Parabolic Press, Stanford, 1975).
	
	\bibitem{ZacconePRE2009}
	A.Zaccone, H. Wu, D.Gentili, M. Morbidelli,
	\textit{Theory of activated process under shear with application to shear-induced aggregation of colloids},
	Physical Review E \textbf{80}, 051404 (2009).
	
	\bibitem{Batchelor1976}
	G. K. Batchelor, \textit{Brownian diffusion of particles with hydrodynamic interaction}, Journal of Fluid Mechanics \textbf{74}, 1-29 (1976).
	
	\bibitem{ZacconeNess}
	C. Ness, A. Zaccone, \textit{Effect of hydrodynamic interactions on the lifetime of colloidal bonds}, Industrial \& Engineering Chemical Research, \textbf{56}, 3726-3732 (2017). 
	
	\bibitem{Nosenko}
	V. Nosenko, A.V. Ivlev, G.E. Morfill, \textit{Microstructure of a Liquid Two-Dimensional Dusty Plasma under Shear}, Physical Review Letters \textbf{108}, 135005 (2012).
	
	\bibitem{Murillo}
	T. Ott, M. Bonitz, L. G. Stanton, and M. S. Murillo, \textit{Coupling strength in Coulomb and Yukawa
		one-component plasmas}, Phys. Plasmas 21, 113704 (2014).
	
	\bibitem{Ballauff1}
	M. Fuchs and M. Ballauff, \textit{Flow curves of dense colloidal suspensions: Schematic model analysis of the shear-dependent viscosity near the colloidal glass transition}, J. Chem. Phys. 122, 094707 (2005). 
	
	\bibitem{Naegele}
	A. J. Banchio, J. Bergenholtz, and G. N{\''}agele, \textit{Viscoelasticity and generalized Stokes-Einstein relations of colloidal dispersions}, J. Chem. Phys. 111, 8721–8740 (1999).
	
	\bibitem{Ballauff2}
	M. Ballauff, J. M. Brader, S. U. Egelhaaf, M. Fuchs, J. Horbach, N. Koumakis, M. Krüger, M. Laurati, K. J. Mutch, G. Petekidis, M. Siebenbürger, Th. Voigtmann, and J. Zausch, \textit{Residual stresses in glasses},
	Phys. Rev. Lett. 110, 215701 (2013). 
	
	\bibitem{Ballauff3}
	M. Siebenbürger, M. Ballauff, and Th. Voigtmann, \textit{Creep in Colloidal Glasses},
	Phys. Rev. Lett. 108, 255701 (2012).
	
	\bibitem{Schall}
	Ch. P. Amann, D. Denisov, M. T. Dang, B. Struth, P. Schall, and M. Fuchs, \textit{Shear-induced breaking of cages in colloidal glasses: Scattering experiments and mode coupling theory}, J. Chem. Phys. 143, 034505 (2015).
	
	\bibitem{Likos}
	D. Toneian, C. N Likos and G. Kahl, \textit{Controlled self-aggregation of polymer-based nanoparticles employing shear flow and magnetic fields}, J. Phys.: Condens. Matter 31, 24LT02 (2019).
	
	\bibitem{Melis}
	S. Melis, M. Verduyn, G. Storti, M. Morbidelli, J. Baldyga, \textit{Effect of fluid motion on the aggregation of small particles subject to interaction forces}, AlChe Journal \textbf{45}, 1383-1393 (1999).
	
\end{thebibliography}

\end{document}